# Nanohole etching in AlGaSb with Ga droplets

*Joonas Hilska‡, Abhiroop Chellu‡, Teemu Hakkarainen\**

Optoelectronics Research Centre, Physics Unit, Tampere University, Korkeakoulunkatu 3, 33720 Tampere, Finland.

ABSTRACT

We demonstrate nanohole formation in AlGaSb by Ga droplet etching within a temperature range from 270°C to 500°C, allowing a wide range of tunability of the nanohole density. By leveraging the low vapor pressure of Sb, we can obtain high degree of control over droplet formation and nanohole etching steps and reveal the physics of adatom diffusion in these processes. Furthermore, by combining the experimental results and a geometric diffusion-based model, we can determine the temperature and Sb-flux-dependencies of the critical monolayer coverage of Sb atoms required for driving the droplet etching process to completion. These findings provide new insight into the droplet formation and etching process present in the droplet-mediated synthesis of semiconductor nanostructures and represent a significant step towards development of telecom-emitting quantum dots in the GaSb system.

Semiconductor quantum dots (QD) embedded in a single-crystalline host matrix are important building blocks of emerging quantum technologies based on non-classical light sources, such as

single- or entangled-photon emitters [1,2]. Such quantum nanostructures can be fabricated by several techniques including Stranski-Krastanov growth [3,4], droplet epitaxy [5-7], growth inside pyramidal holes [8], vapor-liquid-solid growth in nanowires [9], and filling of nanoholes formed by local droplet etching (LDE) [10]. The LDE approach is particularly interesting due to its advantages which include narrow exciton linewidths [11], extremely small inhomogeneous broadening [12], bright single-photon emission [13], and vanishing finestructure splitting [14]. These properties have enabled the use of GaAs QDs grown by filling LDE nanoholes in non-classical light sources providing state-of-the-art performance in terms of photon indistinguishability and entanglement [13,15]. However, the operation of the light sources based on LDE is restricted to the 680-780nm spectral range [11] due to the limited direct bandgap range of (Al)GaAs alloys. While LDE of AlAs, GaAs, and AlGaAs with Al [10], Ga [16-21], and In [22-25] droplets has been investigated in detail and the controllable formation of QDs by filling the nanoholes is well known for the arsenide alloys, the knowledge on droplet etching in the InP and GaSb-based materials is more limited, mostly covering self-running droplets and reactions with In droplets under As flux [26-30]. Furthermore, there is no known methodology for forming strain-free QDs by filling droplet-etched nanoholes in these alloy systems which would provide suitable bandgaps for accessing the technologically important telecom wavelengths [31] and silicon photonic integration [32]. The first step towards this direction is to obtain nanoholes with controllable density and shape.

GaSb-based materials are particularly interesting candidates for quantum photonic applications due to several beneficial properties: (i) the direct bandgap of GaSb is 0.73 eV [33] which is suitable in terms of QD emission at the telecom C-band; (ii) AlGa(As)Sb alloys provide very high refractive index contrast [34] (exceeding that of AlGaAs), which is important for constructing photonic devices; and (iii) the lattice mismatch between GaSb-based materials and dissimilar substrates can be relaxed right at the first interface by formation of a network 90°-dislocations [35,36] and exploitation of nucleation layers [37], which is particularly useful considering direct growth of QD emitters on silicon waveguides for chip-level quantum photonic integration. This technological potential, as well as the general aim for advancing the fundamental understanding of the important metal droplet-mediated processes in semiconductor materials, makes antimonides an extremely interesting subject for studying LDE.

In this letter, we present highly controllable etching of nanoholes in $Al_{0.3}Ga_{0.7}Sb$ surfaces using Ga droplets. We show that LDE can be achieved at the temperature range from 270°C to 500°C, resulting in almost three orders of magnitude change in the nanohole density, and that a remarkable control of the nanohole formation can be achieved by precise calibration of the amount of Sb provided for the LDE process. The low vapor pressure of Sb allows controlling the group V flux deterministically with the needle valve of the cracker source unlike in the case of the arsenide system where low As-fluxes required for LDE are provided by flux switch-off transients and indirect As-fluxes [38]. This enables assessment of the critical amount of Sb required for the nanohole formation process at different temperatures.

The samples were fabricated on n-GaSb(100) substrates using a molecular beam epitaxy (MBE) system equipped with effusion sources for the group III elements while Sb was provided by a valved cracker source. Following a thermal treatment and growth of a GaSb buffer layer, a 100 nm thick $Al_{0.3}Ga_{0.7}Sb$ layer was grown at 500°C using Ga and Al fluxes of $J_{Ga}$=0.7 monolayers/s (ML/s) and $J_{Al}$=0.3 ML/s, respectively. The samples were then set to target pyrometer temperatures between 270-500°C for the Ga droplet deposition and LDE. The droplets were formed by Ga deposition with $J_{Ga}$=0.7 ML/s. For both the Ga deposition and LDE, the Sb flux $J_{Sb}$ was set to a predetermined value ranging from 0 to 0.067 ML/s by carefully adjusting the Sb-valve opening. $J_{Sb}$ was calibrated to true atomic flux by depositing a bulk Sb film on GaSb at low temperature (<40°C) and calculating the equivalent $J_{Sb}$ based on the film thickness and deposition time assuming unity sticking coefficient. All fluxes and deposited coverages are defined in ML/s and ML, respectively, where the Sb coverage $\theta_{Sb}$=1 ML corresponds to the number of Sb atoms required for the formation of 1 ML of stoichiometric GaSb, and similarly for the Ga coverage $\theta_{Ga}$.

The LDE begins with the formation of droplets on the AlGaSb surface by Ga deposition. Typically, this step is carried out in a small group V background [38]. In our case, the same value of $J_{Sb}$, set by the needle valve, was used for droplet formation and for the subsequent etching step. The droplet formation is preceded by saturation of the surface reconstruction with Ga atoms [5], thus the droplet formation begins after $\theta_{Ga}$ exceeds a critical coverage $\theta_{c,Ga}$. The droplet growth is terminated by switching off the Ga flux. The sample is subsequently annealed in a small group V flux, which in case of III-As materials is provided by the residual As-flux remaining in the MBE system after closing the As needle valve [38]. In case of Sb, we can control these small fluxes using the needle valve because of the low vapor pressure of Sb and the lack of indirect Sb fluxes. This enables precise and repeatable control of the group V flux which plays an important role in LDE. The Sb atoms from the vapor phase are dissolved in the Ga droplet, which causes nucleation at the triple-phase line (TPL) where the vapor, liquid, and solid phases meet. During this kinetic process, Sb atoms from the solid material below the droplet are dissolved in the liquid, which enables further nucleation. As a result of the TPL nucleation, the droplet size decreases, and at the same time it drills a hole in the semiconductor surface and creates a ring of solid material which is formed by the Ga atoms from the droplet together with the Sb from the vapor phase and from the substrate. The second role of Sb is to form an Sb-terminated surface surrounding the droplets.

The resulting gradient in the surface chemical potential acts as a driving force for the diffusion of Ga atoms away from the droplet and causes layer-by-layer growth of GaSb within the diffusion length of Ga around the droplet. These steps of the LDE process are presented in Fig. 4 (a) and discussed in detail in the text.

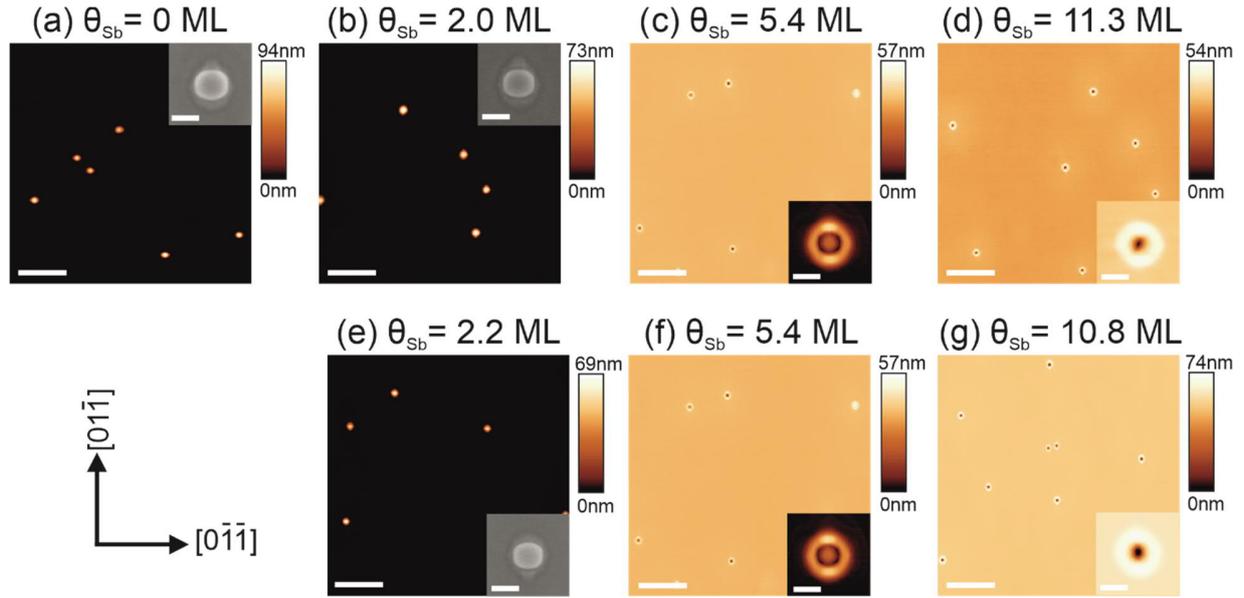

**Figure 1.** Atomic force microscope (AFM) images showing the effect of Sb coverage $\theta_{Sb}$ on the nanohole etching process at 500°C after deposition of 3.2 ML of Ga. In (a)-(d), the etching time is kept constant at 180 s and $\theta_{Sb}$ is adjusted by setting the Sb-flux $J_{Sb}$ to 0 ML/s, 0.011 ML/s, 0.030 ML/s, and 0.063 ML/s, respectively. In (e)-(g), $J_{Sb}$=0.030 ML/s and $\theta_{Sb}$ is adjusted by setting the etching time to 72 s, 180 s, and 360 s, respectively. The scale bars in (a)-(g) are 2 µm. The insets in (a), (b), (e) show scanning electron microscopy (SEM) images and in (e)-(g) AFM close-ups of typical droplets/nanoholes. The scale bars in the insets are 200 nm.

Figure 1 presents the AlGaSb surface morphologies after droplet formation and LDE at 500°C. From Fig. 1(a)-(d), it is evident that the total amount of Sb atoms impinging the surface during the LDE process is a crucial parameter. With $\theta_{Sb}$=0 (Fig. 1(a)), we observe metallic Ga droplets which have just slightly reacted with the solid surface during the 180 s annealing time. With $\theta_{Sb}$=2.0 ML (Fig. 1(b)), the metallic droplets remain, but their height is reduced significantly. At $\theta_{Sb}$=5.4 ML (Fig. 1(c)) some of the holes are completely etched while some still contain a small liquid droplet. At this point the ring caused by the TPL nucleation is already clearly visible. Finally, at $\theta_{Sb}$=11.3 ML, all holes are completely etched and there is no sign of liquid Ga. The final nanohole morphologies are similar to what has been observed for (Al)GaAs etching with Ga and Al droplets [10,16-19]. Similar intermediate and final morphologies are observed also in Fig. 1(e)-(g), where $\theta_{Sb}$ is adjusted by changing the annealing time with constant $J_{Sb}$. There is clearly a critical value for $\theta_{Sb}$ that is needed for consuming the liquid droplet and completing the LDE process. In case of Fig. 1, it is approximately 5.4 ML.

Figure 2 presents the results of LDE at different temperatures, showing that nanoholes can be successfully formed by LDE throughout the temperature range from 270°C to 500°C with some important temperature dependencies in the morphology. The resulting morphologies for temperatures ranging from 353°C to 500°C are nanoholes surrounded by a single ring and a disc. However, concentric rings are observed at the lowest temperature of 270°C. These ring structures are similar to the ones formed during Ga droplet crystallization on AlGaAs by controlling the diffusion process with As [20,21,39]. It is also evident from Fig. 2 that the density increases with decreasing temperature as expected from the thermally activated diffusion during droplet formation [5]. Consequently, the deposited 3.2 ML of Ga is divided between a larger number of droplets, thus causing a decrease of the hole size and depth as a result of a reduction in the droplet volume. Furthermore, the size of the disc formed by layer-by-layer growth around the droplet decreases as the temperature is decreased. Just like the droplet formation, the diffusion of Ga atoms away from the droplet during LDE is a thermally activated diffusion process, but it should be noted that the energetics of these two processes are different due to the differences in the surface chemistry.

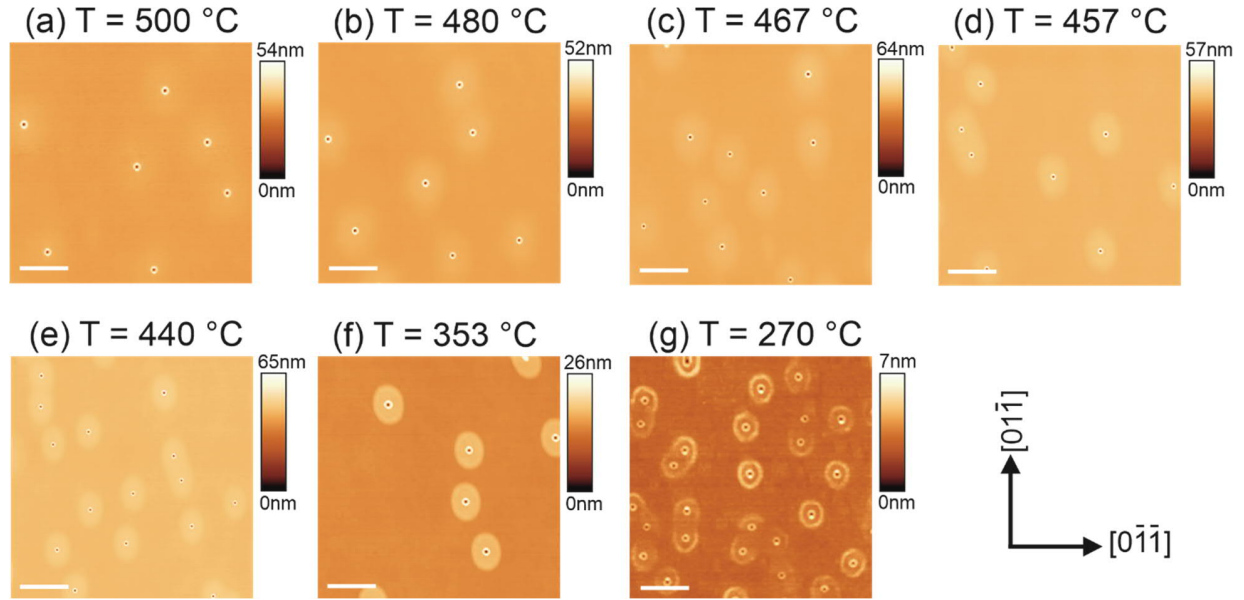

**Figure 2.** AFM images of the AlGaSb surface after deposition of 3.2 ML of Ga for droplet formation and 180 s annealing with $J_{Sb}$=0.060 ML/s at different temperatures. The scale bars are 2 µm in (a)-(e), 600 nm in (f) and 200 nm in (g).

The density of the nanoholes shown in Fig. 2 is plotted as a function of growth temperature in Fig. 3, which shows nearly 3 orders of magnitude increase in the density when the temperature is deceased from 500°C to 270°C, ranging from ultra-low densities in the $10^6$ cm$^{-2}$ range to $2\times10^9$ cm$^{-2}$. The measured density values exhibit linear behavior on semilogarithmic scale for the higher temperatures, while the density value of the two lowest growth temperatures deviate from this trend. According to the phase diagram of GaSb surface by Bracker *et al*. [40], the surface reconstruction of a GaSb surface is (1x3) in the high temperatures but changes to (2x5) when the temperature is decreased, with the transition temperature depending on the Sb beam pressure. As explained in the Supporting Information (SI), we find that $\theta_{c,Ga}$=1.15 ML for the growth at 500°C, while at 270°C, $\theta_{c,Ga}$=1.6 ML, which is consistent with the coverages required for saturating (1x3) and (2x5) reconstructions, respectively, with Ga adatoms [41]. With $J_{Sb}$=0.060 ML/s which was used for the samples presented in Fig. 2 and Fig. 3, the critical transition temperature is expected to be around 390°C [40]. Therefore, the density vs. temperature behavior with a transition region at around 400°C can be explained by the change of surface reconstruction which influences the adatom diffusivity. It should also be noted that coarsening effects by Ostwald ripening, which is typical for the high-temperature LDE of AlGaAs [42], are not observed in Fig. 3. For temperatures above 400°C, we can model the droplet (and nanohole) density using the scaling law [43]

$$N(T) = N_0 e^{E_{A1}/k_B T} \qquad (1)$$

where $N_0$ is a pre-exponential factor, $E_{A1}$ is the activation energy, and $k_B$ is the Boltzmann constant. By fitting Eq. (1) to the experimental data in Fig. 3, we obtain $N_0$=2.1×10$^3$ cm$^{-2}$ and $E_{A1}$=0.51 eV. This is well in agreement with the activation energy of 0.54 eV reported for the Ga droplets on AlGaAs [42].

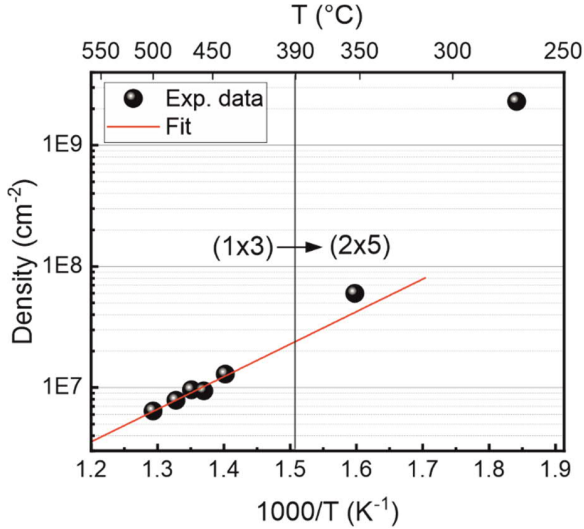

**Figure 3.** Droplet density as a function of growth temperature. The experimental data points are from Fig. 2. The vertical dashed line presents the temperature at which the surface reconstruction

of GaSb is expected to change from (1x3) to (2x5) when $J_{Sb}$=0.060 ML/s [40]. The solid line is a fit of Eq. (1) to the experimental data points in the (1x3) reconstruction regime.

The LDE process takes place after the droplet formation in the presence of a small Sb flux as depicted in Fig. 4(a). Sb atoms from vapor phase dissolve in the liquid Ga droplet and increase the Sb concentration towards supersaturation. In typical droplet epitaxy conditions used e.g. for QD growth, the group V flux enables complete crystallization of the droplet at its initial location [5]. In LDE conditions, which usually involve higher temperatures and/or lower group V fluxes, the decomposition of the solid material underneath the droplet provides another group V source into the droplet and causes the formation of the nanohole. The ring structure surrounding the hole is formed as a result of nucleation when Sb atoms originating either from the vapor phase or from the solid material below the droplet crystallize as GaSb at the TPL as a result of increased Sb concentration in the droplet, with the solubility of Sb in liquid Ga being just 0.3% at 400°C [44]. Some group III and group V atoms can also be lost by evaporation to the vapor phase, particularly at the higher end of LDE temperatures. These nucleation and decomposition processes play an important role in the formation of the nanoholes and the surrounding ring structures, as described by the kinetic models [42,45], but also diffusion processes take place during LDE. Just like the droplet formation (presented in Fig. 4(a) panel 1), the layer-by-layer growth around the nanohole is also a thermally activated process driven by adatom diffusion. It happens during LDE (Fig. 4(a) panel 2) as Sb atoms arriving on the AlGaSb surface from the vapor phase form an Sb-rich surface which changes the surface reconstruction and creates a chemical potential gradient which drives Ga diffusion away from the droplet. The Ga atoms stack with the Sb atoms in a layer-by-layer manner within the diffusion length and form the elliptical discs. However, the conditions during the layer-by-layer growth around the nanohole are more Sb-rich than during the Ga droplet formation, which takes place on a Ga-saturated reconstruction. These differences in the surface chemistry affect the Ga diffusivity. Consequently, the dimensions of the ellipses are significantly smaller than the distance between the droplets. Following the procedure from [46], the Ga diffusion length $l$ during the layer-by-layer growth can be obtained from the disc dimensions as

$$\Delta R = R_2 - R_1 = l = \sqrt{D_{Ga}\tau} = \sqrt{D_0 e^{-E_{A2}/k_B T} \frac{N_S}{J_{Sb}}} \qquad (2)$$

where $R_1$ and $R_2$ are the radii of the ring structure and the disc, respectively, as illustrated in Fig. 4(a), $D_{Ga}$ is the Ga diffusion constant and $\tau$ is the adatom lifetime on the surface. $D_{Ga}$ can be expressed by the exponential diffusion equation where $D_0$ is the diffusivity prefactor and $E_{A2}$ is the activation energy. The adatom lifetime $\tau$ is obtained from $J_{Sb}$ and the surface site density $N_s=5.4\times10^{14}$ cm$^{-2}$. The disc structures shown in Fig. 2 are elliptical, and thus the diffusivity should be examined separately for the $[0\bar{1}1]$ and $[0\bar{1}\bar{1}]$ crystal directions.

The values of $\Delta R_{[0\bar{1}1]}$ and $\Delta R_{[0\bar{1}\bar{1}]}$ were obtained from several nanoholes for each temperature by measuring values of $R_{2[0\bar{1}1]} - R_{1[0\bar{1}1]}$ and $R_{2[0\bar{1}\bar{1}]} - R_{1[0\bar{1}\bar{1}]}$ from cross-sectional AFM profiles (Fig. 4(b)). In case of the highest growth temperatures where the disk is very diffused, $R_{2[0\bar{1}1]}$ and $R_{2[0\bar{1}\bar{1}]}$ were estimated from the AFM maps (Fig. 4(d)) which provided more reliable data.

By fitting Eq. (2) to the experimentally obtained values of $\Delta R_{[0\bar{1}1]}$ and $\Delta R_{[0\bar{1}\bar{1}]}$ (Fig. 4(d)), we obtain the parameters of the physical diffusion process of Ga atoms along the orthogonal $[0\bar{1}1]$ and $[0\bar{1}\bar{1}]$ directions. As a result, we get $E_{A2}=1.05+/-0.010$ eV and $D_0=0.019\times(1.2^{+/-1})$ cm$^2$/s for the $[0\bar{1}1]$ direction, and $E_{A2}=0.98+/-0.011$ eV and $D_0=0.003\times(1.21^{+/-1})$ cm$^2$/s for the $[0\bar{1}\bar{1}]$ direction. These activation energies are smaller than the value of 1.31+/-0.15 eV reported for the Ga LDE of GaAs [46], as expected from the generally higher temperatures needed for LDE of (Al)GaAs, and the different bond dissociation energies of Ga-Sb (192 kJ/mol) and Ga-As (210 kJ/mol) [47].

As shown in Fig. S4 in Supp. Info, the final depth of the nanoholes formed in AlGaSb by LDE with Ga droplet scales linearly with the lateral dimension $R_1$ within the investigated range of growth parameters. Therefore, the hole depth can be controlled by the selection of growth parameters, namely the temperature and $\theta_{Ga}$, during the droplet formation stage.

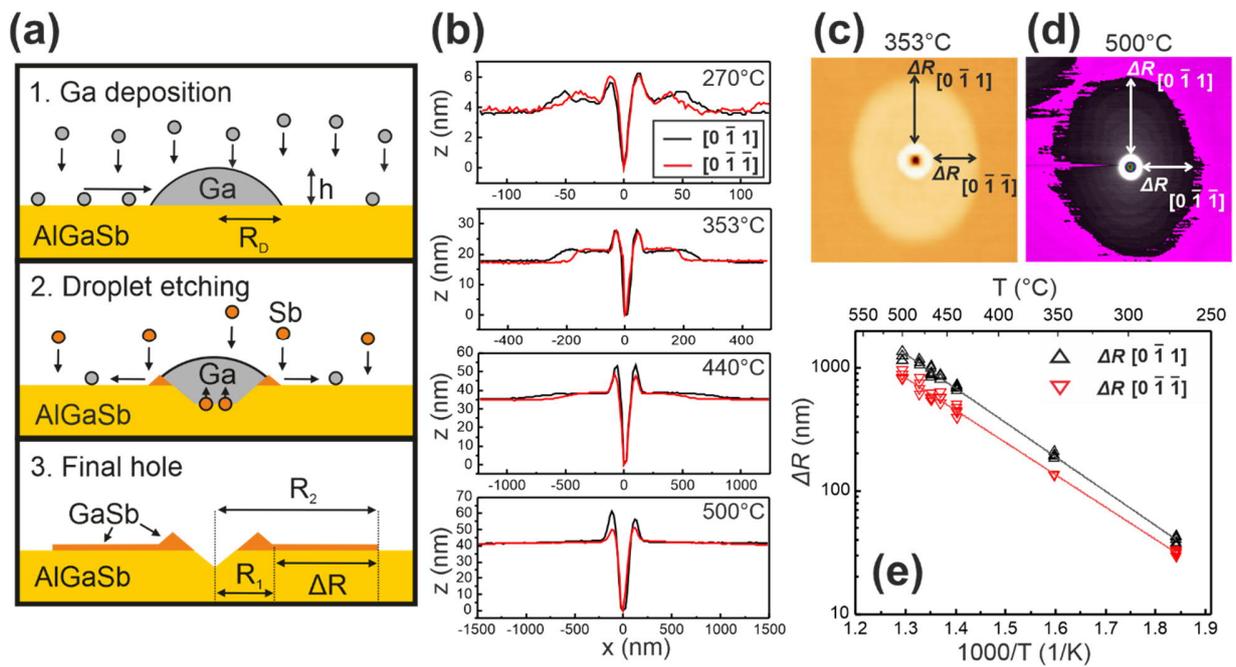

**Figure 4.** (a) An illustration of the phases of droplet formation and droplet etching. The critical dimensions of the droplets and final hole morphologies are indicated. (b) Cross-sectional profiles of the holes and surrounding disc structures etched at different temperatures. The anisotropy of the layer-by-layer growth process is accounted by analyzing $\Delta R = R_2 - R_1$ in the orthogonal $[0\bar{1}1]$ and $[0\bar{1}\bar{1}]$ directions, as shown in (c) and (d). (e) shows the experimental data of $\Delta R$ for $[0\bar{1}1]$ and $[0\bar{1}\bar{1}]$ directions and fits according to Eq. (2).

To our knowledge, no previously reported data is available for the adatom diffusivity on (Al)GaSb. However, rectangular mound defects consisting of spiral step edges on GaSb surface are elongated in the $[0\bar{1}1]$ direction [48] which is consistent with our results, revealing longer diffusion length in the $[0\bar{1}1]$ direction than in the $[01\bar{1}]$ direction.

Now that the temperature dependencies of the diffusion processes are known for both droplet formation and etching, we can analyze in more detail the critical Sb coverage $\theta_{c,Sb}$ needed for completing the LDE process. Assuming unity sticking coefficient and no diffusion for Sb, we can formulate $\theta_{c,Sb}$ from the fact that during LDE, the number of Sb atoms $n_{Sb}$ arriving from vapor phase within the Ga diffusion length from the droplet, i.e. on the elliptical area defined by $R_{2[0\bar{1}1]}$ and $R_{2[01\bar{1}]}$, should be equal to $n_{Ga}$, the initial amount of Ga atoms in the droplet. When this condition is satisfied, the droplet is completely consumed either to nucleation at the TPL (direct impingement of Sb to the droplet) or to the layer-by-layer growth within the Ga diffusion length. By accounting also for the desorption of Ga and Sb atoms during the LDE process, this condition can be formulated as

$$n_{Ga} = \frac{1}{2}\frac{\theta_{Ga}-\theta_{c,Ga}}{N(T)} \times N_{GaSb} = \frac{1}{2}\pi R_{2[0\bar{1}1]}R_{2[01\bar{1}]}\theta_{c,Sb} \times N_{GaSb}\frac{1-P_{Ga}}{1-P_{Sb}} = n_{Sb}\frac{1-P_{Ga}}{1-P_{Sb}}, \quad (3)$$

where $N(T)$ is the droplet density from Eq. (1), and $N_{GaSb}$ is the atom density in GaSb. $P_{Ga}$ and $P_{Sb}$ are the probabilities for the evaporation of Ga and Sb atoms, respectively, back to the vapor phase during the LDE process. In case of congruent evaporation, $P_{Sb} = P_{Ga}$ and the evaporation probabilities cancel out from the balance equation. The factors of ½ on both sides are due to the definition used for ML coverages of Ga and Sb. By replacing $R_2$ with $R_1+\Delta R$ and solving for $\theta_{c,Sb}$ we get

$$\theta_{c,Sb} = \frac{1-P_{Ga}}{1-P_{Sb}} \times \frac{\theta_{Ga}-\theta_{c,Ga}}{N(T)\pi(\Delta R_{[0\bar{1}1]}+R_{1[0\bar{1}1]})(\Delta R_{[01\bar{1}]}+R_{1[01\bar{1}]})}. \quad (4)$$

Since the outer edge of the ring surrounding the etched nanohole is defined by the nucleation at the TPL right in the beginning of the LDE process, the radius of the ring $R_1$ can be replaced with the initial droplet radius $R_D$, as illustrated in Fig. 4(a). The droplet shape can be described as a spherical cap with a volume $V = \frac{1}{2}\pi\kappa R_D^3\left(1 + \kappa^2/3\right)$, where $\kappa = \frac{h}{R_D}$ and $h$ is the initial droplet height. From the AFM profiles of droplets grown at 500°C, we estimate that $\kappa$=0.46+/-0.023 (see Fig. S1 in the SI), which can be assumed to be independent of $V$ and $T$ as long as $R_D$>20 nm [49]. The initial number of Ga atoms in the droplet can thus be expressed as

$$n_{Ga} = \frac{V\rho_{Ga}}{m_{Ga}} = \frac{1}{2}\pi\kappa R_D^3\left(1 + \kappa^2/3\right)\frac{\rho_{Ga}}{m_{Ga}} = \frac{1}{2}\frac{\theta_{Ga}-\theta_{c,Ga}}{N(T)} \times N_{GaSb}, \qquad (5)$$

where $\rho_{Ga}$=5.9 g/cm³ and $m_{Ga}$=69.7 u are the density and atomic mass of Ga, respectively. By solving Eq. (5) for $R_D$, we get

$$R_D = \left[\frac{\theta_{Ga}-\theta_{c,Ga}}{\pi\kappa\left(1+\kappa^2/3\right)N(T)}\frac{m_{Ga}}{\rho_{Ga}} \times N_{GaSb}\right]^{\frac{1}{3}}. \qquad (6)$$

Now Eq. (4) can be rewritten as

$$\theta_{c,Sb} = \frac{1-P_{Ga}}{1-P_{Sb}} \times \frac{\theta_{Ga}-\theta_{c,Ga}}{N(T)\pi(\Delta R_{[0\bar{1}1]}+R_D)(\Delta R_{[0\bar{1}\bar{1}]}+R_D)}, \qquad (7)$$

which includes temperature-dependent expressions for $N(T)$ and $\Delta R_{[0\bar{1}1],[0\bar{1}\bar{1}]}$ from Eq. (1) and Eq. (2) with the fitted pre-exponential factors and activation energies (Fig. 3 and Fig. 4), respectively.

Figure 5 shows $\theta_{c,Sb}$ plotted from Eq. (7) as a function of growth temperature for different values of $J_{Sb}$ assuming $P_{Sb} = P_{Ga}$. This diffusion-based model predicts that $\theta_{c,Sb}$ decreases as temperature is increased and increases when the $J_{Sb}$ is increased. Both effects are consequences of the change in the Ga diffusivity. $J_{Sb}$ reduces the adatom diffusivity (Eq. (2)) and the temperature increases it. The diffusion length during LDE affects the size of the elliptical area which participates to the two processes that consume the Ga droplet: formation of the ring structure by nucleation at the TPL and layer-by-layer growth driven by the Ga diffusion away from the droplet. The temperature dependency of $\theta_{c,Sb}$ is particularly interesting since one might expect that less Sb is required for consuming the small droplets formed at low temperatures. However, the effect of the reduction of the thermally activated Ga diffusion during LDE is significantly stronger than that of the reduction of the droplet volume. The model is also consistent with the experimental findings presented in Fig. 1(a)-(c), which show that, for $J_{Sb}$=0.030 ML/s, $\theta_{c,Sb}$ should be close to 5.4 ML since in Fig. 1(b) we observe some completely etched holes and some holes which still contain a small liquid droplet. Furthermore, the temperature-dependency predicted by the model is in agreement with our experimental findings. For $J_{Sb}$=0.060 ML/s (Fig. 2), we find that with $\theta_{Sb}$=10.8 ML all holes are completely etched when T=500°C, while for T=353°C we find that 19% of the holes still contain some liquid Ga (See Fig. S5 and S6 in the SI), which confirms that more Sb is required for completing the etching process at a lower temperature. The model slightly underestimates the $\theta_{c,Sb}$ at 500°C which can be explained by the increase of Sb evaporation to the vapor phase. This growth temperature is just above the congruent evaporation point of GaSb [50], beyond which the assumption $P_{Sb} = P_{Ga}$ is not precisely valid as $P_{Sb}$ increases faster than $P_{Ga}$ as a function of temperature.

The understanding of $\theta_{c,Sb}$ is important for controlling LDE in order to provide enough Sb for completing the nanohole etching process in the given set of growth conditions, but still avoiding unnecessarily long etching. Furthermore, by selecting appropriate Sb fluxes and etching times, it is possible to adjust the Ga diffusion during LDE while still ensuring completion of the LDE process.

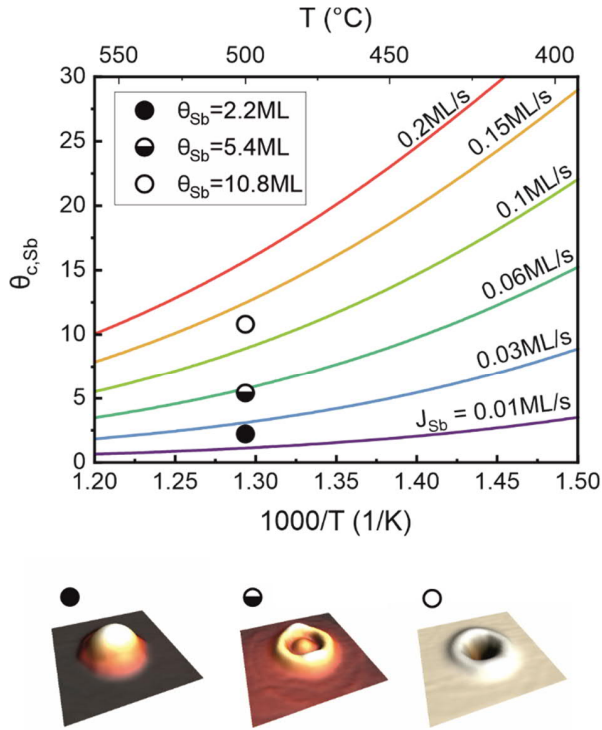

**Figure 5.** Critical Sb coverage calculated for different Sb fluxes $J_{Sb}$ from Eq. (7) for $\theta_{Ga} - \theta_{c,Ga}$=1.93 ML, which corresponds to deposition of 3.2 ML of Ga on a (1x3) reconstructed surface, while taking into account the Ga atoms consumed in planar growth due to reaction with the Sb atoms (as explained in the SI). The experimental data points from Fig. 1 (e)-(g) $J_{Sb}$=0.030 ML/s are presented by the round symbols. The filled, half-filled, and empty symbols correspond to the Sb coverage that results in incomplete etching with liquid droplets remaining, situation close to the critical Sb coverage with liquid Ga remaining only in some of the droplets, and completely

etched holes with no liquid Ga present, respectively. These phases of the etching process are presented in the 3D AFM profiles showing droplet and nanohole morphologies.

In conclusion, we demonstrated highly controllable formation of nanoholes in $Al_{0.3}Ga_{0.7}Sb$ by LDE with Ga droplets over a temperature range from 270°C to 500°C, which provides tunability of the nanohole density by almost three orders of magnitude. The low vapor pressure of Sb enables precise control of the Sb flux, which provides high degree of repeatability and provides means for adjusting the diffusion processes in LDE. We have analyzed the diffusion of Ga adatoms in both droplet formation and LDE steps and presented a model for predicting the critical amount of Sb required for complete hole etching and consumption of the liquid Ga droplets. This work represents a significant advance towards LDE-based quantum confined structures in GaSb-based materials which provide direct bandgaps covering the important wavelength range from telecom to mid-infrared.

ASSOCIATED CONTENT

**Supporting Information**. Determination of the droplet shape; Critical Ga coverage for droplet formation; Nanohole depth; Assessment of the completion of the nanohole etching process.

AUTHOR INFORMATION

**Corresponding Author**

*teemu.hakkarainen@tuni.fi

**Author Contributions**

The manuscript was written through contributions of all authors. All authors have given approval to the final version of the manuscript. ‡These authors contributed equally.

**Funding Sources**

Academy of Finland Project QuantSi (decision No. 323989)

Academy of Finland Project NanoLight (decision No. 310985)

ACKNOWLEDGMENT

The authors acknowledge financial support from the Academy of Finland Projects QuantSi (decision No. 323989) and NanoLight (decision No. 310985)